\documentclass[aoas,nameyear,MSNbibl,seceqn,dvips]{arximspdf}
\usepackage{graphicx}

\doi{10.1214/10-AOAS438}
\volume{5}
\issue{2B}
\pubyear{2011}
\firstpage{1360}
\lastpage{1378}

\makeatletter
\newcommand{\Ac}{\mathcal{A}}
\renewcommand{\mid}{|}

\def\@bmisc[#1]{%
  \get@battribute{unstr}%
  \common@pub@types%
  \let\bauthor\bbl@bauthor%
  \let\bhowpublished\@firstofone%
  \def\borganization##1{{\bauthor@style ##1}}%
}

\makeatother

\begin{document}
\begin{frontmatter}

\title{False discovery rates in somatic mutation studies of~cancer\thanksref{TT1}}
\runtitle{False discovery rates in somatic mutation studies}

\begin{aug}
\author[A]{\fnms{Lorenzo} \snm{Trippa}\corref{}\ead[label=e1]{ltrippa@jimmy.harvard.edu}}
\and
\author[A]{\fnms{Giovanni} \snm{Parmigiani}\ead[label=e2]{gp@jimmy.harvard.edu}}
\runauthor{L. Trippa and G. Parmigiani}
\affiliation{Dana-Farber Cancer Institute and Harvard School of Public Health}
\address[A]{Dana-Farber Cancer Institute\\
3 Blackfan Circle, Boston\\
USA\\
and\\
Harvard School of Public Health\\
677 Huntington Avenue\\
 Boston, Massachusetts 02115\\
 USA\\
\printead{e1}\\
\hphantom{\textsc{E-mail}: }\printead*{e2}} %adresu isvedimo komanda gale!
\end{aug}
\thankstext{TT1}{Supported in part by NSF Grant DMS-03-42111.}

% HISTORY:
\received{\smonth{5} \syear{2010}}
\revised{\smonth{10} \syear{2010}}

% ABSTRACT
%
\begin{abstract}
The purpose of cancer genome sequencing studies is to determine the
nature and types
of alterations present in a typical cancer and to discover genes
mutated at high frequencies.
In this article we discuss statistical methods for the analysis of
somatic mutation frequency
data generated
in these studies. We place special emphasis on a two-stage study design
introduced
by Sj{\"o}blom et~al. [\textit{Science} \textbf{314} (2006) 268--274].
In this context, we describe and compare statistical
methods for constructing
scores that can be used to prioritize candidate genes for further
investigation and to
assess the statistical significance of the candidates thus identified.
Controversy has surrounded the reliability of the false discovery rates
estimates provided by the approximations used in early cancer genome studies.
%We provide a comprehensive assessment of several proposals using a
%large collection of highly realistic simulated data sets.
To address these, we develop a semiparametric Bayesian model that
provides an accurate fit to the data. We use this model to generate a
large collection of realistic scenarios, and evaluate alternative
approaches on this collection.
Our assessment is impartial in that the model used for generating data
is not used by any of the approaches compared.
And is objective, in that the scenarios are generated by a model that
fits data.
%We discuss a model based approach for comparing statistical methods
%for controlling false discovery rates.
Our results quantify the conservative control of the false discovery
rate with the Benjamini and Hockberg method compared to the empirical
Bayes approach and
the multiple testing method proposed in Storey
[\textit{J. R. Stat. Soc. Ser. B Stat. Methodol.}
\textbf{64} (2002) 479--498].
Simulation results also show a negligible departure from the target
false discovery rate for the methodology used in Sj{\"o}blom et~al.
 [\textit{Science}
\textbf{314} (2006) 268--274].
\end{abstract}

% KEYWORDS
%

\end{frontmatter}

%s1 ###
\section{Introduction}\label{sec1}
The systematic investigation of the genomes of human cancers has
recently become possible with
improvements in sequencing and bioinformatic technologies. \citet{Sj07}
and \citet{Wo07}
determined the
sequence of
%a set of 13,023
comprehensive collections of coding
genes % termed the consensus coding sequences
(CCDS and RefSeq) in colorectal and
breast cancers, and provided a catalogue of somatic mutations. In this
context, a somatic mutation
is a tumor-specific mutation not present in the germline of the patient
whose tumor contained
it. Subsequently,
\citet{Gr06}
investigated somatic mutations in the coding exons of 518
protein kinase genes in a large and diverse set of human cancers. More
recently, mutation data from glioblastoma tissues have been studied in
the \citet{At08} and \citet{Pa08},
and from pancreatic cancer in \citet{Jones2008p1065}.
%Other projects are under way or
%planned, including the Cancer Genome Atlas Project
%(http://cancergenome.nih.gov/), whose
%goal is to identify all the genes mutated in at least 5\% of human
%cancers, across cancer types.
Statistical analysis of the data generated in these studies poses new
challenges that are worthy
of careful consideration.
\citet{Gr06}
have provided an in-depth analysis of data generated
by one-stage studies. To make optimal use of sequencing resources,
\citet{Sj07}
introduced a two-stage
design, with the stages termed ``Discovery'' and ``Validation.'' The
Discovery Stage consists of a
catalog of mutations in all genes considered, for example, all genes in
the CCDS database. This design
permitted selection of the subset of genes that harbored at least one
somatic mutation, termed
``Discovered.'' This subset was further investigated in a Validation
Stage which cataloged somatic
mutations in discovered genes in an independent set of tumor samples.
Genes that were mutated
in at least one tumor in the Validation set were termed ``Validated.''
In this article we consider this two-stage design.
\citet{Sj07} and \citet{Wo07},
adopting this experimental design, discovered that among the genes
whose mutations are likely responsible for carcinogenesis, the
majority, the ``hills,'' had
mutations in small subgroups of cases, while only a handful of genes,
the ``mountains,''
were mutated in large subgroups. %The Wood et al. study \cite{Wo07}
Thus, the hills and not the mountains dominate the cancer genome landscape.
This imbalance emphasizes the importance of statistical methods
in identifying the mutations involved in the carcinogenesis process.

The somatic mutations found in cancer tissues are either ``drivers'' or
``passengers'' [\citet{Wo07}]. Driver mutations
are causally involved in the neoplastic process and are positively
selected during tumorigenesis.
Passenger mutations provide no positive or negative selective advantage
to the tumor but are
retained by chance during repeated rounds of cell division and clonal
expansion. The overarching
goal of the statistical analysis of cancer mutation data is to identify
genes that are most likely to contain driver mutations
on the basis of their mutation type and frequency. This is done by
quantifying the evidence that
the mutations in a gene reflect underlying mutation rates that are
higher than the passenger rates.

Early cancer genome projects provided a rank order of genes by their
potential to be drivers of carcinogenesis based on mutation frequencies
in tumors as well as the
genes' size and nucleotide compositions.
%likelihood that each gene may
%harbor passenger mutations.
To provide an indication of the significance of lists of possible
driver genes, they also provided estimates of the false discovery rate
% stat background & unsolved problems
%The statistical analysis of cancer sequencing data is primarily
%focused on identifying cancer genes.
%This includes the application of testing methods suitable for the
%so-called "large $p$ small $n$" paradigm;
% a small number of cancer tissues are used for simultaneously testing
%a large number of hypothesis, one for each sequenced gene.
%Benjamini and Hokberg \cite{BH95}
%introduced a practical multiple hypothesis testing procedure which
%controls the false discovery rate
(FDR), that is, the expected proportion of putative drivers that are
actually passengers, or the proportion of erroneously rejected null
hypotheses [\citet{BH95}].
%During the last decade it has been recognized by a relevant body of
%literature that, in many applications, it is appropriate controlling
%the FDR.
Since the seminal article of \citet{BH95}, several authors have proposed
alternative methods that control the FDR with improved operating
characteristics.
%A relevant part of them
%building on results in Benjamini and Hochberg \cite{BH95}.
Important contributions
include
% have been given for
relaxing the independence assumption among the test statistics
%used in Benjamini and Hochberg \cite{BH95}.
and handling discrete statistics
%Control of the FDR when test statistics are discrete has also been
%studied
[\citet{By01}].
%Some proposals have been studied for taking into account \emph{a
%priori} beliefs.
Reviews
% on multiple testing methods which account for the main contributions
%in multiple testing methods
are given in
\citet{Du03}, \citet{Ch07} and \citet{Fa08}.
Despite a~rich literature,
%on statistical methods for controlling FDR,
applications often include heuristic arguments and, in most cases, the choice
of a method is far from trivial: a trade-off arises between methods
which have been proved through rigorous analytical arguments to control
the FDR under a specified threshold and less conservative procedures
that rely on heuristic arguments, examples and asymptotic theory.
The trade-off between methodological rigor and operating
characteristics becomes particularly relevant, for example,
when the independence hypothesis is inappropriate or when data are
highly discrete.
% Little is known about how to efficiently take into account the
%dependency between test statistics and, as a result,
Applications often ignore
these problems.

% our contribution
In this article we consider methods that are specific to somatic
mutation analysis, and assess them
%assess alternative methods
by a novel model-based approach.
% reproducing in silico the experiment by means of a model based
%approach is discussed.
Our idea is to develop a ``super partes'' model that % is neutral
%considered methods a
can be used to provide highly realistic artificial datasets for method
evaluation. We refer to these as data-driven simulated scenarios.
Our %The main contribution consists in proposing a simple
scheme for evaluating alternative methods
% the proposed scheme can easily
can take into account possible
discrepancies between methods' assumptions and the data. %, by
%including them in a Bayesian model.
%dependency structures of the test statistics, as indicated by the
%data, by including them in a Bayesian model.
%Our proposal is motivated by
We apply this comparison method by revisiting the \citet{Sj07} methodology
%; the sequencing data has been analyzed adopting alternative
%approaches,
%see
as well as the alternative approaches proposed shortly afterward by
several groups [\citet{Fo07}, \citet{Ge07}, \citet{Ru07}
and
\citeauthor{Pa07}  (\citeyear{Pa07,GP2007})].
We discuss the application of this scheme to the data collected in \citet
{Wo07}. %The Wood it al. \cite{Wo07} study extends the Sjoblom et al.
%The next paragraph emphasizes some peculiarities of the study. An
%advantage of of our approach is that it requires to formally specify
%the investigator assumptions by means of a Bayesian model. The
%probability model can %then be used as a comparison tools for
%alternative methods.

% brief description of Wood et. al study

% Our strategy

%In this article a comparison of alternative methodological approaches
%for assessing which genes constitute a positive selection factor
%and causes carcinogenesis is given. The illustrated comparisons are
%based on data driven simulations. A genuine Bayesian approach is
%adopted for generating synthetic scenarios
%consistent with the data available from the study discussed in Wood et
%al. \cite{Wo07}. Such an approach constitutes a simple framework for
%generating scenarios with strong similarities with the data.

% article structure

The article is structured in 5 sections. In Section~\ref{sec2} we introduce the
notation for modeling mutation counts in tumor tissues and present our
probability model. In Section~\ref{sec3}
we briefly review techniques for controlling or estimating false
discovery rates, and specific approaches % discussed in
%Forrest and Cavet \cite{Fo07}, Getz et al. \cite{Ge07}, Rubin and
%Green \cite{Ru07} and Parmigiani et al. \cite{Pa07}
for the analysis of somatic mutation data.
% use the probability model for generating possible scenarios and
%comparing alternative methods for the analysis of sequencing data.
In Section~\ref{sec4} we compare these approaches.
%In Section~\ref{sec5} the data collected in the Wood el al. study are analyzed
%using the compared methods in order to confirm the simulation study
%findings.
Final remarks are given in Section~\ref{sec5}.

%s2 ###
\section{Data-driven simulation scenarios}\label{sec2}
\label{probmod}
%patient X gene X type model
%s2.1 ###
\subsection{Approach}\label{sec2.1}
The idea of data-driven simulations is structured in two steps.
First we select a flexible probability model for the observed
mutations. The model includes model-specific parameters, genes-specific
latent variables and observable mutation counts.
%Our primary focus is on the genes-specific latent variables which
%discriminate drivers from passengers.
The model is consistent with widely established assumptions on
carcinogenesis. % and includes prior distributions on the unknown
%parameters.
Prior distributions on the unknown model parameters are specified by
using both external experimental results and heuristic data-driven approaches.
Second, we infer the genes-specific latent variables, including driver
\mbox{status}, through a Monte Carlo Markov chain (MCMC) algorithm and
generate simulated data sets consistent with the
%synthetically reproducing the experiment discussed in
\citet{Wo07} study.
%Specifically, we approximately sample from the conditional
%distribution of the latent variables and match in silico the
%experimental data adopting a predictive approach. In principle this
%approach allows
%us to simulate experiments assuming alternative experimental designs
%and constitutes a solid tool for comparing both alternative methods
%for the data analysis as well as alternative designs.
%end{itemize}

Throughout the article this Bayesian model is used only for generating
simulated data sets that are highly consistent with the observed data
in \citet{Wo07} and are impartial
to the FDR approaches examined. In principle, the model could be used
directly for selecting
genes having high posterior probabilities of being drivers.
Other potential applications of the model include (i) predicting the
experimental
outcomes for ongoing studies, (ii) optimally choosing the resource
allocation for the validation stage using a decision theoretic approach
and, more generally, (iii) developing adaptive strategies
for sequencing experiments.
%Nevertheless, here
Nevertheless, here we prefer limiting our comparisons to highly
computationally efficient methods whose operating characteristics can
be assessed via Monte Carlo techniques over a large number of data sets.
Moreover, it would be circular to simultaneously use our Bayesian model
for FDR estimation and for generating data for evaluation.
%Assessing alternative statistical methods for the analysis of cancer
%genome sequencing data is particularly useful:
As noted in \citet{Ge07}, the performance of some methods could be
sensitive to the distribution underlying the experimental data; %
%rigorous comparisons under
using a~probability model that fits the observed data, but also
averages across plausible values of the unknown parameters,
%and addresses the relevant uncertainties
allows us to provide appropriately objective comparisons.

%%%%%%%%%%%%%%%%%A
%
%t1 ###
\begin{table}[b]
\tabcolsep=0pt
%\tablewidth=310pt
\caption{Summary of notation for the data produced by the
study for the $g$th gene, and~associated~gene-specific parameters}
\label{tab1}\begin{tabular*}{\textwidth}{@{\extracolsep{\fill}}l@{\hspace*{-0.25cm}}l@{}}
\hline
\multicolumn{2}{@{}l@{}}{Mutation counts}  \\
 \quad $X_{gm}^1$
& number of mutations of type $m$ detected in gene $g$ in the
Discovery Stage.
\\[1pt]
 \quad $X_{gm}^2$
& number of mutations of type $m$ detected in gene $g$ in the
Validation Stage.
\\
[5pt]
\multicolumn{2}{@{}l@{}}{Coverage}  \\  \quad $T_{gm}^1$ & coverage of type $ m$
in gene $g$ in the Discovery Stage. \\
 \quad $T_{gm}^2$ & coverage of type $m$
in gene $g$ in the Validation Stage. \\
[5pt]
Mutation rates &\\  \quad $\gamma_m^1$
& rate of mutation of
type $m$ in the Discovery Stage.
\\
 \quad $\gamma_m^2$
& rate of mutation of
type $m$ in the Validation Stage. \\
 \quad $\theta_g$ & multiplicative gene-specific random effect. \\
\hline
\end{tabular*}
\end{table}

%In this section we propose a Bayesian model for driver and passenger
%mutations. We will use the model as an "independent ombudsman" for
%comparing
%selection procedures finalized to identifying driver genes. Other
%potential applications of the model includes (i) prediction of the
%experimental
%outcomes for ongoing studies, (ii) optimal choice of the resource
%allocations for the validation stage using a decision theoretic
%approach and, more generally, (iii) developing adaptive strategies
%for sequencing experiments, though these are not further explored here.
%s2.2 ###
\subsection{Sampling model}\label{sec2.2}
While cancer genome sequencing projects produce a wealth of
information, in this article we will focus on the somatic mutation
counts, broken down by gene and context, and
considered separately for the Discovery and Validation Stages. Table~\ref{tab1}
summarizes the notation
we will use. Our model applies to a single disease, say, colorectal
cancer, at a~time.
The probability model is defined on the basis of a few well-established
assumptions that allow to specify a distribution of mutation counts
conditional on unknown gene-specific mutation rates.
%We start from the evidence that mutations in cancer genes
%increase the risk of developing the disease. It is therefore natural
%assuming a joint model which accounts for the mutations and the
%possibility of developing the disease.
%The data available for cancer sequencing studies usually refer only to
%cohort of patients, there is usually no control. It is therefore
%convenient a model such that
%the distribution of the mutations conditionally on the disease status
%is analytically tractable.
%The use of
Using latent variables, we
%allows us to
specify a model % consistent with the above considerations.
allowing for unknown composition of the genome in terms of passengers
and drivers, and for heterogeneous mutation rates across driver genes.
% We also allow for tumors to have different mutation rates.
More formally, we assume that, for each gene and sample, the number of
mutations of type $m$, that is, the number of identical mutations that
can occur only in a specific context, for example, C to G in a CpG
locus, is a~mixture of Poisson distributions,
%
%e2.1 ###
\begin{equation}\label{mod1}
X_{igm}\mid\theta_{g},\rho_i , \eta_m \sim \operatorname{Poisson} ( \rho_i \eta_m
\theta_{g} T_{igm} ) \quad\mbox{and}\quad\theta_{g}
\stackrel{\mathrm{i.i.d.}}{\sim} F,
\end{equation}
where $i$ indexes a tumor, $g$ indexes a gene and $T_{igm}$ denotes the
coverage, or the number of successfully sequenced nucleotides
susceptible to mutation of type $m$ in gene $g$ and sample $i$. The
parameter $ \eta_m $ represents the rate of mutations of type $m$ among
passengers.
The multiplicative factor $\rho_i $
is designed to capture the fact that the abundance of mutations varies
across tumors.
The transition of a tissue from normal to cancer can be described as a
progressive accumulation of mutations, some of them are drivers and
some are passengers.
This dynamic varies across tumors;
such heterogeneity is the focus of a significant portion of cancer
research; see \citet{St09} for a stimulating discussion.
Our model, as well as those used in the previously mentioned cancer
studies, describes a snapshot of this dynamic process at the time of sequencing.
The product $ ( \rho_i \eta_m) $ can be interpreted as the rate of
mutations of type $m$ in tumor $i$, assuming that the nucleotide is
part of a passenger gene.
The gene-specific latent variables~$\theta_g$ capture gene-specific
variation across the genome; if $\theta_g=1$, gene $g$ is a~passenger,
while higher values identify the drivers.
Our model assumes that the rates of
mutation across different types $m$ of mutations in a single gene $g$
are proportional to
the rates of mutation in a passenger gene.
Finally, $F$ is the distribution of $\theta_g$'s across the genome. It
allows for values of $1$ or bigger and allows for a concentration of
mass on the value of~$1$, corresponding to the passengers.

The overall structure of the model reflects the assumption that the
drivers have higher rates of mutation than the passengers. The analyses
in \citet{Sj07} and \citet{Wo07}
were aimed at selecting cancer genes with mutation rates higher than
the hypothesized passenger rates.
% noncoding loci; possible mutations in noncoding regions do not
%constitute a positive selection factor in carcinogenesis and can
%be used for comparisons with putative cancer genes.
The use of the Poisson distribution in (\ref{mod1}) is motivated by the
fact that, under mild assumptions, it well approximates a more rigorous
multinomial model [\citet{Wo07}] obtained by modeling possible mutations
in a single gene as binary variables. % for rare events, in our case
%the mutations.

% gene X type model

Values of individual passenger rates %The investigator can
%approximately measures the intensities
$ ( \rho_i \eta_m) $ can be obtained using data external to the somatic
mutations counts [\citet{Sj07}, \citet{Wo07}].
%One strategy for measuring passenger rates consists in sequencing
%; we refer to Sjoblom et al. \cite{Sj07} for a detailed account on
%the biological arguments that allows
%the investigator to estimate these quantities. Such estimates are
%based on sequencing data which refer to genome loci other than those
%considered in the analysis.
%In synthesis, from a statistical point of view, it is relevant
%that the measurements of
% The intensities $ ( \rho_i \eta_m) $ can be measured on the basis of
%external experimental data that are not used for identifying cancer
%genes.
This allows us to simplify the model (\ref{mod1}) by collapsing data
across patients. %Our primary interest is on the latent variables $
Once the intensities $ ( \rho_i \eta_m) $ are known,
the collapsed counts data $(X_{gm}=\sum_{i} X_{igm})$
are sufficient statistics for evaluating the likelihood function. %A
%more practical argument for collapsing the mutations counts is the
This allows for a considerable reduction of the computational
requirements. % complexity necessary for identifying cancer genes.
We will thus use the alternative representation of (\ref{mod1}):
%
%e2.2 ###
\begin{equation}\label{mod2}
X_{gm}\mid\theta_{g}, \gamma_m \sim \operatorname{Poisson} ( \gamma_m \theta_{g}
T_{gm} ) \quad\mbox{and}\quad\theta_{g}  \stackrel{\mathrm{i.i.d.}}{\sim} F,
\end{equation}
where $ X_{gm} $ is the total number of mutations of type $m$ harbored
in the $g$th gene, $T_{gm} =\sum_i T_{igm} $ and $T_{gm} \gamma_m=\sum
_i \rho_i \eta_m T_{igm}$.

% two phase model

Multistage designs are attractive strategies for identifying cancer
genes; the first stage indicates a subset of genes that are
more likely to be drivers % whose mutations are likely of
%being involved in carcinogenesis
and, in the subsequent phase, only this subset is analyzed.
Relevant cost--effectiveness analysis and comparisons of alternative
designs for genome-wide studies are illustrated in
\citet{Sa02}, \citet{Sa03}, \citet{Sa04}, \citet{Kr06}, \citet{Sk06}, \citet
{Wa06} and \citet{Pa09}.
In the studies discussed by \citet{Sj07} and \citet{Wo07}, at the end of
the first
stage all genes which harbored one or more mutations are considered for
further study. We will use the notation $X_{gm}^1$ and $X_{gm}^2$ for
denoting the number of mutations in the two phases. Similarly,
$T_{gm}^1$ and $T_{gm}^2$ denote the coverages and $(\gamma_m^1, \gamma
_m^2)$ the rates for the discovery and validation phases. Model (\ref
{mod2}) can be adapted to the two-stage design as follows:
%
%e2.3 ###
\begin{eqnarray}\label{mod3}
  X_{gm}^1\mid\theta_{g}, \gamma_m^1 &\sim& \operatorname{Poisson} ( \gamma_m^1 \theta
_{g} T_{gm}^1 ),
\nonumber\\
  X_{gm}^2\mid X_{gm}^1, \theta_{g}, \gamma_m^1&\sim&
\cases{\displaystyle  0 ,&\quad  if  $X_{gm}^1=0$, \cr\displaystyle
\operatorname{Poisson}( \gamma_m^2
\theta_{g} T_{gm}^2) ,&\quad  if $ X_{gm}^1>0$,
}
\\
 \theta_{g}  &\stackrel{\mathrm{i.i.d.}}{\sim}& F . \nonumber
\end{eqnarray}
In these expressions the coverages $T_{gm}^1$ and $T_{gm}^2$ are
considered fixed. While some variation may be experimentally observed,
this is unlikely to be related to a gene's driver status, and, thus,
it is appropriate to model the data conditionally on the coverages.
For our purpose, the following three considerations are critical:

\textit{Two-stage design.} Only genes
that harbor at least one mutation in the Discovery Stage are sequenced
in the Validation Stage. This screening condition needs to be taken
into account when assessing significance using $p$-values or other
methods that rely on the sampling
distribution.% of the data, as it removes a large number of possible
%experimental outcomes.

\textit{Coverage.} The number of nucleotides successfully sequenced
is generally smaller than the gene length times the number of tumors
analyzed. For example,
certain exons may be technically challenging to sequence. It is
appropriate to apply stringent
quality criteria to sequencing data, which lead to the exclusion of
nucleotides whose sequence could
not be identified with certainty. Nucleotides excluded, or not covered,
%% are not eligible for false positive mutations and
should not be included in statistical evaluations.
%Moreover, when applying corrections
%for multiple testing, coverage should be a consideration for all the
%genes, including those for which
%no mutations were found.
In what follows $p$-values and other statistics used for prioritizing
putative driver genes are computed
taking into account, for each gene, which loci have been sequenced.

\textit{Context.}
The third consideration involves the
observed bases in the mutations. In sequencing studies, the precise
bases that comprise the mutation,
as well as the neighboring bases, termed ``mutation contexts,'' are
important.
We use a classification of contexts provided in \citet{Wo07}.
This is a partition of the sequenced basis. For our purpose, it is
relevant that each subset has specific rates of mutation under the
passenger status.
%
%For our purpose these are relevant because different contexts
%We assume that the context of a passenger locus determines
%It is well known that the sequence of basis of a specific locus
This fact implies that the priority given to a
gene in further studies and the statistical significance of a gene
should depend not only on the number of nucleotides
but also on the gene-specific basis sequence.

%s2.3 ###
\subsection{A Bayesian approach to generating simulation scenarios}\label{sec2.3}
% prior
For our analysis we propose embedding the sampling model (\ref{mod3})
in a Bayesian semiparametric model. This will allow us to generate
possible scenarios consistent with the data from the \citet{Wo07} study.
We use a Dirichlet process prior [\citet{Fe73}] for the unknown
distribution $F$,
%
%e2.4 ###
\begin{equation}\label{modd2}
F\sim \operatorname{Dirichlet} (\Ac),
\end{equation}
where $\Ac$ is a positive measure on $[1,\infty)$.
Reviews on the Dirichlet process and applications in biostatistics are
given in \citet{Dunson} and \citet{Mueller}.

% Why Dirichlet?

%Our Bayesian semiparametric model assumes that the rates of
%mutation across different types $m$ of mutations in a single cancer
%gene $g$ are proportional to
%the rates of mutation in a passenger gene.
%Possible extensions of the Bayesian model are beyond the scope of the
%present paper.
%Extensions, which relax the semi-parametric assumption, and
%comparisons with alternative models are given Ding and Parmigiani
%(2010).
%Ding and Parmigiani (2010) also elaborate on the idea of using an
%informative hyper-prior, on the parameters $\gamma_m$, centered
%on
%estimates obtained using sequencing data of noncoding loci.
%Here we only emphasizes the
We chose the Dirichlet mixture model because of its flexibility
compared to alternative parametric prior distributions. Simple
preliminary analysis of the
mutation data shows that the so-called mountains can have mutation
rates over 100-folds higher than the passengers, while hills have
markedly lower rates. %These genes, usually called mountains, in most
%cases, were already
%identified in the literature as cancer genes.
%On the other hand, the rates of the majority of other likely driver
%genes are remarkably lower. These are the hills.
% The clear distinction between mountains and hills
%has been reported in Wood et al. \cite{Wo07}.
To capture both the mountains and the hills, we specify a sufficiently
flexible prior on the unknown mixing distribution $F$.
As shown in \citet{Vent}, Bayesian mixtures model effectively heavy tail
distributions; in contrast, the posterior behavior
of more parsimonious parametric models can be strongly biased.

% MCMC

In order to generate scenarios consistent with the data in \citet
{Wo07}, we use a Monte Carlo Markov chain algorithm
% similar to
%the Gibbs sampler
discussed in \citet{Es95}.
The sampler is based on the Polya urn representation of the Dirichlet
process [\citet{Bl73}] and
has been studied for posterior simulation under the generic Dirichlet
mixture model
\[
p(X_1,\ldots ,X_n \mid F)= \prod_{i=1}^n \int p(X_i\mid\theta)\,dF(\theta),
\qquad F\sim \operatorname{Dirichlet}(\Ac).
\]
The only condition for implementing the sampling scheme of \citet{Es95}
is that for every subset $\{i_1,\ldots ,i_m\}$ of distinct integers
ranging from $1$ to $n$, the integral
%
%e2.5 ###
\begin{equation}\label{integ}
\int\prod_{j=1}^m p(X_{i_j}\mid\theta) \,d\Ac(\theta)
\end{equation}
can be easily computed. To this end, we specify the prior parameter $\Ac
$ proportional to a spiked distribution, including
a point mass at $\{1\}$ and a~shifted gamma distribution with support
$[1,\infty)$, that is,
%
%e2.6 ###
\begin{equation}\label{DG}
\Ac(dx)\propto \delta_1(dx)+ c I(x\ge1) e^{-a(x-1)} (x-1)^{b-1}\,dx,
\end{equation}
where $ \delta_1(\cdot)$ is a Dirac measure, $I(\cdot)$ is the
indicator function and $a,b,c$ are strictly positive.
It can be verified that this choice of $\Ac$ allows us to analytically
solve integral (\ref{integ}).

% heuristic elicitation

We chose the mean of the random distribution $F$ by a simple procedure.
We recall that the centering distribution of the Dirichlet process is
$\Ac(dx)/\Ac([1,\infty))$; for every subset $B$ of the real line, if
$0<\Ac(B)<\Ac([1,\infty))$, then $F(B)$,
a priori, is beta distributed, with mean $\Ac(B)/\Ac([1,\infty))$;
see \citet{Fe73}.
%We specify the centering distribution
In order to specify the centering distribution, we first compute the
maximum likelihood estimator $\hat F$ of the mixing distribution in
(\ref{mod3}).
Then, we specify the parameter $\Ac(\{1\})/\Ac([1,\infty))$, this is
the a priori expectation of the unknown proportion of passenger genes.
%The maximum likelihood estimator $\hat F$
In what follows $\Ac(\{1\})/\Ac([1,\infty))$ is set equal to $\hat F
([1,2) )$.
%Sensitivity to implemented posterior inference also sensitivity issues
% Here the value of 2 is chosen heuristically.
Finally, the parameterization of the centering distribution is
completed by setting $a$ and $b$
in (\ref{DG}) so that the means and variances of the two distributions
$\hat F(dx)$ and $\Ac(dx)/\Ac([1,\infty))$ are identical.
The steps outlined have clear interpretations, nevertheless,
the choice $\Ac(\{1\})/\Ac([1,\infty))=\hat F ([1,2) )$ is, to some
extent, arbitrary; we also implemented posterior inference for
alternative prior parameterizations:
these include $\Ac(\{1\})/\Ac([1,\infty))$ equal to $\hat F ([1,1.5)
)$, $\hat F ([1,3) )$ and $\hat F ([1,4) )$. We did not observe marked
sensitivity; for example, the ratio between the maximum and the minimum
posterior estimates of the number of drivers is equal to
1.06.

We use the contexts and mutations classification discussed in \citet{Wo07}.
This classification is important because it allows us to account for
variations in the rates of mutations for passengers across loci, with
rates depending on the basis sequences.
This classification includes 25 possible types of mutations ($m=1,\ldots
,25$). The rates for the 1st and the 2nd stage $(\gamma_m^1 \mbox{ and }
\gamma_m^2)$
were measured using SNP data.
The SNP-based approach estimates the passenger mutation rates by
comparing the nonsynonymous
to synonymous mutation ratios in cancer and normal tissues.
This approach has been considered in \citet{Wo07}. It estimates the
passenger rates using sequencing data from loci
which are known for not contributing to carcinogenesis, and thus
are not positively selected during carcinogenesis; this characteristic
is the defining feature of passenger genes.
%Note that this peculiarity defines the passenger status.
% we don't like mle

A key advantage of the Bayesian estimate of the mixing distribution,
compared to the maximum likelihood
estimate $\hat F$, is that it allows us to % incorporate an \emph{a
%priori} guess on $F(\{1\})$, the unknown proportion of passenger genes.
%In contrast $\hat F(\{1\})=0$ almost surely when the $(T_1^0,T_1^1,
%Moreover the Bayesian model, when used for sampling scenarios, allows
%us
fully take into account the uncertainty on the distribution of rates
across genome, and to produce data sets under many different plausible versions
of this distribution.

% producing scenarios

%f1 ###
\begin{figure}[b]

\includegraphics{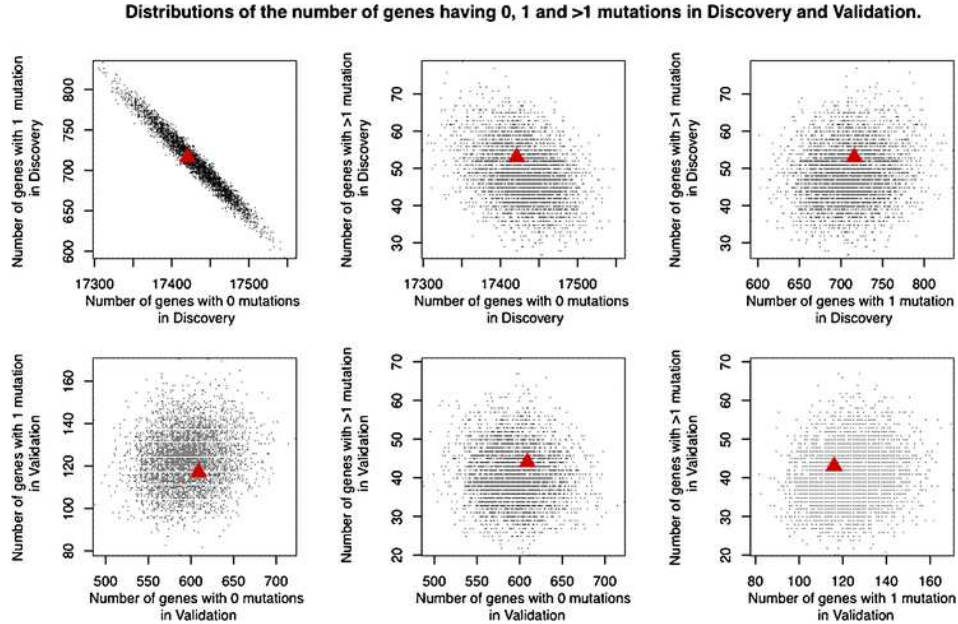}

\caption{Correspondence between simulated and observed counts of
mutated genes in 10,000 simulated scenarios.}
\label{fig1}\end{figure}

The MCMC algorithm outlined, after a sufficient number of burn-in
iterations, produces approximate samples from the conditional
distribution of
the latent variables given the data. The number of burn-in iterations
can be assessed by means of standard diagnostic procedures for MCMC
methods; see, for example, \citet{Sm07}.
%reviews alternative diagnostic tools and illustrates a software for
%implementations.
Each iteration of the MCMC algorithm provides a collection of $\theta
_g$'s which is used as a simulation scenario.
% us to sample approximately uncorrelated samples from
%$p(\theta\mid X)$.
For each scenario we generate a single data set $X$ using (\ref{mod3}).
%Once the experimental design $(T^1,T^2)$ and the latent variables $
%completed by generating the data $X$ from (\ref{mod3}).
Each scenario can be used to evaluate a given list of putative drivers
by checking the proportion of genes in the list for which $\theta_g=1$.
% Figure~\ref{fig1}

To highlight the excellent fit we obtain, Figure~\ref{fig1} provides an
overview %synthetic representation
of 10,000 scenarios sampled by means of the proposed approach. Each
iteration attempts to reproduce
the experiment on colorectal tumors discussed in \citet{Wo07}. They
considered 18,190 genes.
During the discovery phase, all genes were sequenced in 11 tumors. %
%were sequenced and,
During the validation phase,
769 genes were sequenced in 24 tumors. In the discovery phase, 17,421
genes did not harbor any mutation, 716 genes harbored 1 mutation and 53
genes harbored more than 1 mutation.
In the validation phase, 609 did not harbor any mutation, 115 harbored
1 and 45 harbored more than 1. Figure~\ref{fig1} represents the distribution,
across the simulated scenarios,
of the number of genes harboring 0, 1 or more than 1 mutations in the
two stages.
As appropriate, scenarios vary in their number of mutated genes found,
while distributions are centered around the observed data.

%%%%%%%%%%%%%%%%%%%%%%%

The data-driven simulation approach %for assessing alternative methods
%designed for prioritizing gens
requires the variability across simulations to be consistent with the
data.
In principle, if the experiment was repeated several times, one would
like to observe a similar degree of variability across simulations and
across experiments.
In practice, the degree of variability across simulations can be
critically evaluated by means of inferential arguments.
The \mbox{bootstrap} method is specifically designed for predicting, on the
basis of a single experiment, the degree of variability across
independent replicates of the experiment.
We compared the degree of variability across simulations, illustrated
in Figure~\ref{fig1}, with the variability estimates obtained by bootstrapping.
Our application of the bootstrap builds on
parametric estimates, for each gene, of the probabilities $p(\sum_m
X_{gm}^1=1)$, $p(\sum_m X_{gm}^1> 1)$, $p(\sum_m X_{gm}^1\ge1, \sum_m
X_{gm}^2=1)$ and $p(\sum_m X_{gm}^1\ge1 ,\sum_m X_{gm}^2> 1)$, obtained
by fitting logistic binary regression models; these estimates are
functions of the gene-specific coverages $T_{gm}^1$. The degree of
uncertainty represented in Figure~\ref{fig1} agrees with the bootstrap
estimates; the ratios between the standard deviations of the six
univariate empirical distributions illustrated in Figure~\ref{fig1} and the
corresponding bootstrap estimates range between 1.03 and~1.17.
Note that the six variables considered in Figure~\ref{fig1} define a coarse
partition of the genes; more generally, one can use the bootstrap
method for assessing the degree of variability of other marginal distributions.

%%%%%%%%%%%%%%%%%%%%%%%%%%

%s3 ###
\section{Alternative methods for controlling the FDR}\label{sec3}

In this section we review alternative FDR methods, we will then compare
their operating characteristics by means of the simulations described
in the previous section.

%s3.1 ###
\subsection{The Benjamini and Hochberg, and the Storey procedures}\label{sec3.1}

%BH setting

Benjamini and Hochberg (\citeyear{BH95}) considered which of the null hypotheses
$(H_g, g=1,\ldots,\break G)$, if any, should be rejected given $p$-values $(Z_g,
g=1,\ldots,G)$, one for each hypothesis.
They proposed a procedure for rejecting a (possibly empty) subset of
hypotheses so as to control the FDR, that is,
%
%e3.1 ###
\begin{equation}\label{FDRD}
\mathbb{E} \biggl( \frac{\mathrm{number\ of\ erroneously\ rejected\ hypotheses}}
{\mathrm{number\ of\ rejected\ hypotheses}}\biggr),
\end{equation}
with the proviso that the above ratio is $0$ when none of the
hypotheses is rejected. The expectation in (\ref{FDRD}) is with respect
to the unknown joint distribution of the test statistics $Z_g$.
The input of the procedure is the vector $(Z_g, g=1,\ldots ,G)$ and the
output is the subset of rejected hypotheses.
%A critical assumption is that
The $ p$-values corresponding to the true null hypothesis are
independently uniformly distributed.
% Both these conditions are violeted in somatic mutation studies.
The FDR is an attractive error measurement in many applications with
massive multiple hypotheses testing; we refer to \citet{Du03} for a
comparison with alternative
error measurements.

%BH theorem

Let $(Z_{(1)},\ldots,  Z_{(G)})$ be the sorted values in ascending order
of the $p$-values $(Z_{1},\ldots,  Z_{G})$ and let $\alpha\in(0,1)$ be
any desired upper bound for the FDR.
\citet{BH95} proved that the procedure that rejects the hypotheses with
a $p$-value lower than
%
%e3.2 ###
\begin{equation}\label{BH}
\max\biggl\{ \{0\} \cup\biggl \{ Z_{(g)} \dvtx  Z_{(g)}< \alpha \frac{g}{G} \biggr\} \biggr\}
\end{equation}
controls the FDR below $\alpha$. They show that, for every hypothetical
proportion $p_0$ of the true null hypothesis,
%
%e3.3 ###
\begin{equation}\label{ineq}
\operatorname{FDR}\le p_0\alpha.
\end{equation}
The above inequality shows that the procedure is conservative. \citet
{St02} studied an alternative step-up method which starts from
an estimate $\hat p_0$ of the proportion of the true null hypothesis
and then set a threshold similar to~(\ref{BH}) indicating the rejection region.
The estimate $\hat p_0$ and the inequality~(\ref{ineq}) are used for
inflating the upper bounds $\alpha(g/G)$, $g=1,\ldots ,G$, in~(\ref{BH}).
The estimate of $p_0$ is based on the right tail of the empirical
distribution of the $p$-values. % and it allows less conservative
%behavior
%compared to \cite{BH95}.

%s3.2 ###
\subsection{The empirical Bayes method}\label{sec3.2}

% EBM assumptions and goal

The use of empirical Bayes procedures for estimating the FDR has been
discussed by several authors, including \citet{Ef01}, \citet{Ef03} and
\citet{Du08}.
In this approach,
one computes a data summary, or score $Z_{g}$, that captures departure
from the null hypothesis, such
as a $p$-value, a likelihood ratio or other statistics. In our case,
$Z_g$ should capture evidence for
the rate of mutation for gene $g$ being higher than the passenger rate.
The empirical Bayes method is based on a mixture representation of the
distribution of these scores:
%
%e3.4 ###
\begin{equation}\label{EBE}
f(z) = p_0 f_0(z) + (1-p_0) f_1(z) ;
\end{equation}
$f(\cdot)$ is the distribution of a randomly selected score,
$p_0$ is the unknown proportion of true passengers,
$f_0(\cdot)$ is the distribution of a score randomly selected among
passengers and, finally, $f_1(\cdot)$ is the distribution of scores
among drivers.
%Expression (\ref{EBE}) is the core of the empirical Bayes method.
% It is assumed that the statistics $Z_g$ is a realization from the
%mixture distribution (\ref{EBE}).
The objectives are estimating
the conditional probabilities
\[
\frac{ (1-p_0) f_1(Z_g)}{p_0 f_0(Z_g) + (1-p_0) f_1(Z_g)}
\]
and identifying a rejection region $\mathcal{R}$ containing the more
significant scores, such that
%
%e3.5 ###
\begin{equation}\label{selection}
\frac{\int_{\mathcal{R}} p_0 f_0(z)\,dz}{\int_{\mathcal{R}}p_0 f_0(z) +
(1-p_0) f_1(z)\, dz} \le \alpha;
\end{equation}
where $\alpha$ is the target ratio between the mistakenly rejected
hypothesis and the total number of rejections.
%Note that both the above ratios are conditional probabilities.

%EBM strategy

%In order to briefly synthesize the method it is useful, at first
%glance, assuming that $f$ has a discrete support and that lower scores
%$Z_g$ indicate more evidence against the $g$-th hypothesis.
%In such a case,
%The distribution $f$ and the proportion $p_0$ are usually estimated by
%smoothing the scores' empirical distribution.
The distribution $f_0$ can be approximated simulating
the scores assuming the genome only consists of passenger genes [\citet{Wo07}], while the distribution $f$ and the proportion $p_0$ are usually
estimated by smoothing the scores' empirical distribution.
% The proportion $p_0$ can be estimated in several ways depending on
%the details of the application, a possible estimator is
%$\hat p_0= \min(\hat f/f_0) $.
Finally, expression (\ref{selection}) is used for rejecting a subset of
null hypothesis in such a way that the estimated proportion of
erroneously rejected null hypothesis is lower than $\alpha$.
% under a constraint almost identical to (\ref{selection}), i.e. the
%left hand is substituted with
% 1-\frac{\sum_{Z_g\in\mathcal{ R}} \hat p_0 f_0(Z_g)}{\sum_{Z_g\in
%When the statistics $Z_g$ are continuos variables the same approach
%can be adopted; the main difference is that the empirical distribution
%is substituted with a density estimator $\hat f$.

There is an important difference between the \citet{BH95} method and the
empirical Bayes method. The former rejects a subset of a list of null hypotheses
and controls the expected proportion of erroneously rejected
hypotheses. The latter, for a generic rejection region, estimates the
proportion of true null hypotheses; the investigator can then select
a subset of hypothesis such that the estimate is lower than a desired
threshold $\alpha$.

%s3.3 ###
\subsection{The CaMP score}
Sj{\"o}blom et~al. (\citeyear{Sj07}) introduced the cancer mutation prevalence (CaMP) score, to
provide a ranking
of the Validated genes and select promising candidates. The score is
based on the probability of observing the number of actually found mutations
if the gene was a passenger gene, using
% for a passenger gene $g$ when
a binomial model:
%
%e3.6 ###
\begin{equation} \label{multinomial}
 \quad p_g%(X_{g})
=
\cases{
\displaystyle \!\! \prod_m \!\!\pmatrix{T_{gm}^1\cr X_{gm}^1} \!\!( \gamma
_m^1)^{X_{gm}^1}(1- \gamma_m^1)^{T_{gm}^1-X_{gm}^1}  , &  $\displaystyle\sum_m
X_{gm}^1+ X_{gm}^2=0$,\vspace*{2pt}\cr\displaystyle
\!\!\prod_{j=1}^2\prod_m \!\!\pmatrix{T_{gm}^j\cr X_{gm}^j} \!\!(\gamma
_m^j)^{X_{gm}^j}(1- \gamma_m^j)^{T_{gm}^j-X_{gm}^j}
% \binom{T_{gm}^2}{X_{gm}^2} (\gamma_m^2)^{X_{gm}^2}(1-
,&  $\displaystyle\sum_m X_{gm}^1>0$,
\cr\displaystyle
0 ,&\quad   otherwise.
}\hspace*{-25pt}
\end{equation}
Recall that $\gamma_m^1$ and $\gamma_m^2$ are expected proportions of
nucleotides, in a passenger gene,
harboring mutations of type $m$.
%The above expression is the probability that a binomial model assigns
%to the mutations harbored in the $g$-th gene; note that
The model takes into account the two-stage experimental design.
%Then, for every gene, we define $p_g$ as the probability $p( X_{g})$
%of the experimental number of mutations under model (

We then rank the $ p_g$'s and call $q_g$ the resulting ranks. The CaMP
score is defined as
\[
\operatorname{CaMP}_g(X_{g1}^1,\ldots ,X_{gM}^1,X_{g1}^2,\ldots
,X_{gM}^2)=-\infty \qquad\mbox{if } \sum_m X_{gm}^2=0
\]
and
\[
\operatorname{CaMP}_g(X_{g1}^1,\ldots ,X_{gM}^1,X_{g1}^2,\ldots ,X_{gM}^2)= - \log
_{10}(p_g/q_g) \qquad\mbox{if } \sum_m X_{gm}^2>0 .
\]
The top row corresponds to genes that are eliminated at the Discovery
or Validation Stage.
The goal of the CaMP score is to rank genes according to the strength
of the evidence that
they may be mutated at rates higher than the passenger rates.
An advantage of CaMP scores is that they can be easily computed. This
definition %allows an
%heuristic
can also be seen as an approximation of the \citet{BH95} procedure. In
\citet{Sj07} a threshold of 1 on the CaMP scores was considered
to generate a list of putative drivers, with the goal of producing
%The procedure aimed on computing
a list having approximately 10\% of erroneously discovered drivers.
%It has to be noted that
The probabilities $p_g$ are not $p$-values because they are not tail
probabilities, but % The quantities $p_g$
can be used as approximation of the $ p$-values if the expected number
of mutations in each gene is close to $0$.
%Also, in the two stage design, the number of ways in which one may
%observe an outcome more extreme than the actual observation is
%markedly reduced compared to the one--stage case, %reducing the gap
%between point and tail probabilities.

\citet{Fo07} proposed to use tail probabilities for controlling the
FDR, though they use a sampling model that does not account for the
two-stage design.
\citet{Ge07} emphasized that $p$-values can be used for controlling the
FDR and proposed alternative test statistics.
The closer test statistics to the CaMP score are obtained
by computing the distribution of $p_{g}$, under the hypothesis that
the $g$th gene is a passenger gene, and evaluating the resulting tail
probability. This procedure produces $p$-values which can be used for
controlling the FDR by applying the
\citet{BH95} method. \citet{Ge07} noted that the CaMP score is not a
monotone transformation of the probabilities $p_g$; that is, given two
genes $g^{\prime}$ and $g^{\prime\prime}$, it can happen that
$p_{g^{\prime}}<p_{g^{\prime\prime}}$ and $\operatorname{CaMP}_{g^{\prime\prime
}}<\operatorname{CaMP}_{g^{\prime}}$. % They show that the CaMP score can fail
%to correctly prioritize genes likely to be drivers.
\citet{Ge07} also discussed the idea of controlling the FDR by using %
%considered alternative test statistics and proposed to
the log-likelihood ratio, that is,
% Getz et al. considered both the $p$-values defined in Forrest and
%Cavet and the log--likelihood--ratio test statistic
%
\[\log\bigl(p(X_g\mid\theta_g) \bigr)-\log\bigl(p(X_g\mid\hat\theta
_g) \bigr),
\]
where $\theta_g$ represents the hypothesis that gene $g$ is a passenger
and $\hat\theta_g$ is the gene-specific maximum likelihood estimate.
\citet{Ru07} proposed to use the tail probabilities $p (\sum_{m=1}
X_{gm}\ge x )$, based on the aggregate number of mutations in a gene.
%, without consideration of the mutation type $m$.
These critiques of the analysis in \citet{Sj07} and the alternatives
proposed by these authors suggest the idea of systematically assessing
the operative characteristics of alternative approaches.
%We have adopted the Benjamini and Hochberg
Our comparisons in Section~\ref{sec4} consider the
CaMP score and the alternative $p$-values proposed by \citet{Ge07}, \citet
{Fo07} and \citet{Ru07}. % We recall that these $p$-values are the
%likelihood ratio test statistic and the tail probability under the
%null hypothesis.
These alternative statistics are used for implementing both the \citet
{BH95} method, the method discussed in \citet{St02} and the Empirical
Bayes method.
% The Empirical Bayes approach for controlling the FDR requires the
%estimation $\hat p_0$ of true null hypothesis; we estimate this
%proportion by computing the maximum likelihood estimator %for the
%mixing distribution in (\ref{mod3}) while the density $f_0$ is
%approximated
%by simulating the scores under the hypothesis that all genes are
%passengers.

%s4 ###
\section{Simulation study}\label{sec4}
In this section we compare alternative methods for identifying driver
genes using the 10,000 simulation scenarios described in Section~\ref{sec2}.
%As we anticipated our comparison is based on conditional simulations
%of the latent variables in the Bayesian
%model introduced in section (\ref{probmod}). We performed posterior
%simulation by means of the outlined Monte Carlo Markov chain method.
%The number of burn in transitions which allows the Markov chain to
%approximately reach the stationary distribution was chosen using the
%BOA package \cite{Sm07}.
%Then we simulated other 10,000 transitions of the Markov chain and
%selected 250 scenarios, i.e. one realization of the latent variables
%every 40 iterations.
The \citet{Wo07} study considered both colon and breast tumors. %has
%analyzed the mutations across patients with colon cancer and breast
%cancer.
Here we consider each tumor type separately, and repeat the same analysis.
%We consider alternative proposals
%discussed in \cite{Sj07}, \cite{Fo07}, \cite{Ge07}, \cite{Ru07} and
%The parameterization of the prior distribution was chosen following
%the described heuristic approach and allows us to include in the
%probability model
%the external estimates of the rates of mutations for the passenger
%genes.

%
%t2 ###
\begin{table}[b]
\tabcolsep=0pt
\caption{Operating Characteristics of $9$ alternative
procedures. Comparison between the Benjamini and Hochberg (BH), Storey
(ST) and empirical Bayes (EB) methods.
The average operating characteristics have been computed setting the
FDR control at the 10\% and 20\% levels}
\label{sim}
\label{tab2}
\begin{tabular*}{\textwidth}{@{\extracolsep{4in minus 4in}}lccc@{\hspace*{-1pt}}cc@{\hspace*{-1pt}}}
\hline
 &  & \multicolumn{2}{c@{\hspace*{-1pt}}}{\textbf{False discoveries}} & \multicolumn{2}{c@{\hspace*{-1pt}}}{\textbf{Average number of}} \\
&  & \multicolumn{2}{c@{\hspace*{-1pt}}}{\textbf{proportion}} & \multicolumn{2}{c@{\hspace*{-1pt}}}{\textbf{selected genes}}\\[-5pt]
&  \textbf{Scores} & \multicolumn{2}{c}{\hrulefill} & \multicolumn{2}{c@{}}{\hrulefill}\\
\textbf{Method}&\textbf{or $\bolds p$-values} & $\bolds{\alpha=10\%}$ & $\bolds{\alpha=20\%}$
& $\bolds{\alpha=10\%}$ & $\bolds{\alpha=20\%}$ \\
\hline
\multicolumn{6}{@{}c@{}}{Colon-based simulations}\\
BH &
CaMP score & 0.101 & 0.218 & 150.4 & 221.8 \\
BH &
$p(\sum X_{mg}^j > x)$& 0.074 & 0.146 & 115.1 & 198.7 \\
BH & likelihood ratio & 0.071 & 0.144 & 135.2 & 208.2 \\
EB &
CaMP score & 0.106 & 0.232 & 162.3 & 242.6 \\
EB &
$p(\sum X_{mg}^j > x)$& 0.100 & 0.238 & 147.8 & 250.4 \\
EB & likelihood ratio & 0.102 & 0.211 & 163.8 & 237.2\\
ST &
CaMP score & 0.104 & 0.220 & 157.3 & 235.4 \\
ST &
$p(\sum X_{mg}^j > x)$& 0.099 & 0.232 & 146.5 & 242.4 \\
ST & likelihood ratio & 0.100 & 0.207 & 159.0 & 231.6\\
[4pt]
\multicolumn{6}{@{}c@{}}{Breast-based simulations}\\
BH &
CaMP score & 0.098 & 0.196 & 146.6 & 218.8 \\
BH & $p(\sum X_{mg}^j > x)$& 0.075 & 0.150 & 119.4 & 193.7 \\
BH & likelihood ratio & 0.074 & 0.141 & 131.7 & 205.3\\
EB & CaMP score & 0.108 & 0.216 & 158.2 & 226.7 \\
EB & $p(\sum X_{mg}^j > x)$& 0.105 & 0.225 & 139.6 & 235.0 \\
EB & likelihood ratio & 0.098 & 0.209& 153.5 & 223.4\\
ST & CaMP score & 0.103 & 0.213 & 157.6 & 219.5 \\
ST & $p(\sum X_{mg}^j > x)$& 0.101 & 0.222 & 136.8 & 228.6 \\
ST & likelihood ratio & 0.096 & 0.207 & 147.6 & 221.2\\
\hline
\end{tabular*}
\end{table}

Our simulation study compares the performance of alternative methods
for ranking cancer genes and selecting putative cancer genes.
Table~\ref{tab2} provides the average operating characteristics across all the
simulated scenarios.
All methods are set to control the FDR at 10\% and 20\% levels in turn.
%We recall that
The empirical Bayes method estimates the FDR:
%in this case the
the investigator controls the FDR by
approximately matching the desired level $\alpha$ and the estimated
proportion of false discoveries.
% have been used for selecting cancer genes controlling the FDR
%below the 10\%.
Table~\ref{tab2} shows the average proportion of genes erroneously classified as drivers.
%erroneously discovered cancer genes and the average number of selected
%genes.
%
Our results emphasize that operating characteristics are quite
conservative when the FDR is controlled, on the basis of alternative
$p$-values, using
the \citet{BH95} procedure. They also illustrate the importance of
quantifying this conservative behavior by contrasting it with
alternative methods such as the method proposed in \citet{St02} and the
Empirical Bayes method.
%The average operating characteristics allow the investigator both a
%more critic choice of the method and a more comprehensive
%interpretation of the results. % The data produced in the sequencing
%experiment allows us, as shown in Section~\ref{sec2}, to define a probability
%model for simulating the
The Bayesian model of Section~\ref{sec2} allows us to simulate mutations across
the genome,
while the operating characteristics of the alternative methods shown in
Table~\ref{tab2} provide approximations of their performances.
The interpretation of the simulation study is anchored to the modeling
assumptions formalized in Section~\ref{sec2}.
The results provide a solid basis for choosing among alternative methods.

All the methods considered produce lists of putative drivers genes
with an average misclassification error below the 11\% when the control
of the FDR is set at the 10\% level.
If the desired $\alpha$ level is 10 \%, our results indicate that, when
the CaMP scores are adopted, with
the \citet{BH95} procedure, the average proportion of false discoveries
is approximately equal to $\alpha$, while under the empirical Bayes
method and the \citet{St02} method, a slight excess of false discoveries
is observed. We
note, both in the colon and breast simulation studies, that the use of
the likelihood ratios or the $p$-values $p(\sum X_{mg}^j > x)$ with the
\citet{BH95} procedure, on average, selects a substantially
lower number of putative drivers than the alternative approaches. When
the likelihood ratio and the $p$-value $p(\sum X_{mg}^j > x)$ are
compared, under the empirical Bayes method and the \citet{St02} method,
the likelihood ratio seams preferable; the average proportions of false
discoveries are similar, but the likelihood ratio statistics select
larger sets of putative drivers than
the $p$-values $p(\sum X_{mg}^j > x)$.
%We
Also, when comparing the empirical Bayes method and the \citet{St02}
method based on the likelihood ratio statistics, we observed only small
variations
both in the average proportion of false discoveries and in the average
number of putative drivers.
The operating characteristics when $\alpha$ is set at the 20\% level
confirm the conservative behavior of the \citet{BH95} procedure with the
likelihood ratios or the $p$-values $p(\sum X_{mg}^j > x)$, but also show
departures from the target FDR $\alpha$ of the Empirical Bayes method
and the \citet{St02} method.
These results hold for both colorectal and breast cancer. When $\alpha
$ is equal to 20\%, the likelihood ratio seems preferable to the
$p$-value $p(\sum X_{mg}^j > x)$, under both the empirical Bayes method
and the \citet{St02} method; the likelihood ratios select putatitive
drivers with an average misclassification error closer
to the 20\% target than the $p$-values.

\begin{figure}

\includegraphics{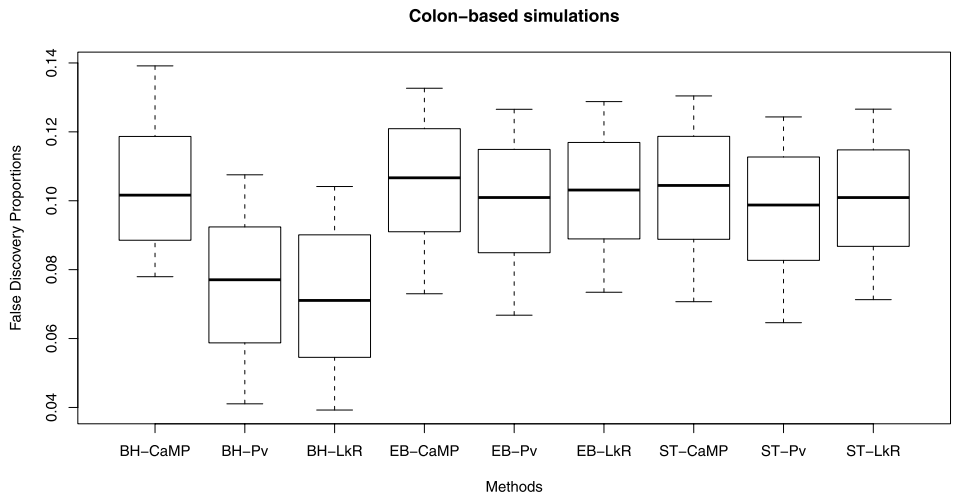}

\caption{The distribution of the true false discovery proportions
across simulated scenarios. Each box plot corresponds to one of the
methods ($\alpha=10 \%$) considered in the simulation study.
The lines of a box plot correspond to the median and to the $10$th,
$25$th, $75$th and $90$th percentiles of the empirical distribution of
the false discoveries proportion.}
\label{var}
\label{fig2}\end{figure}

%The operating characteristics at 20\% FDR confirm the conservative
%behavior of the \cite{BH95} procedure but also show
%departures from the target FDR $\alpha$ of the Empirical Bayes method.
The simulation-based comparison also allows us to asses
the variability of the false discoveries proportion across scenarios,
summarized in Figure~\ref{fig2}. Each box plot is representative of the
simulation-based distribution of the proportion of erroneously rejected
hypothesis. The degree of variability is similar for all the methods
considered; this similarity indicates that the average operating
characteristics concisely reported in
Table~\ref{tab2} are sufficient for a reliable evaluation of the methods.

%
%f2 ###

To illustrate the importance of accounting for
% show that it is relevant that the computation of the tail
%probabilities $P(\sum X_{mg}^j > x)$ takes in to account
the two stages of the experimental design in computing tail
probabilities $p(\sum X_{mg}^j > x)$, we repeated
the analysis ignoring the two-stage structure. That is, we computed the
tail probabilities
under the erroneous assumption that the mutation counts were observed
in a single stage experiment. We observed a substantial reduction of
the average number of selected genes, across simulation scenarios, when
the tail probabilities
are used for implementing the \citet{BH95} procedure; with $\alpha=20\%$
in the colon cancer and breast cancer cases, the averages become $165.6
$ and $157.0, $ respectively, and, with $\alpha=10\%$, they decrease to
$98.4$ and $ 96.2$.
%In contrast, when the tail probabilities are used for implementing the
%Empirical Bayes method negligible variations are observed when the two
%stage design is not taken in to account.

\begin{figure}

\includegraphics{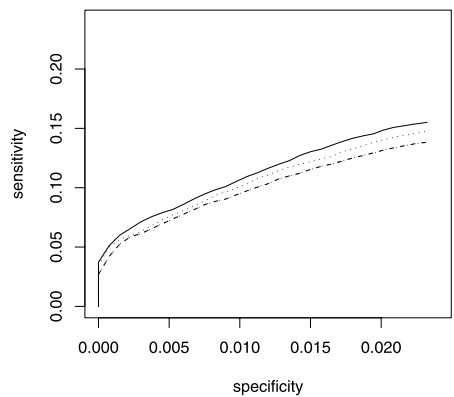}

\caption{Estimated ROC curves for the log-likelihood ratio test
statistics (solid line), the $p$-values $p(\sum X_{mg}^j > x)$ (dotted
line) and the CaMP scores (dashed line).}
\label{var}
\label{fig3}\end{figure}

%We conclude this section with a brief comparison of the CaMP score
%with the $p$-values $P(\sum X_{mg}^j > x)$ and the
%log--likelihood--ratio statistics.
%Our simulation study showed that the log--likelihood--ratio test
%statistics in combination with the Empirical Bayes analysis provides
%minimal improvements in FDR calibration at 10\%, but %more substantial
%improvements at 20\%,
%over
%both the $p$-values $p(\sum X_{mg}^j > x)$ and the CaMP scores.
Another important question concerns the fidelity of the ranking
provided by these statistics. Figure~\ref{fig3} shows the Bayesian estimates of
the left side of the ROC
curves corresponding to each of the scores considered.
The ROC curves are computed by separately estimating the scores'
distributions across drivers and across passengers.
The partial area under the curve is truncated at the 2\% specificity level.
This is equal to
% (corresponding to the log--likelihood--ratio test statistic) is
$2.05\times10^{-3}$ for the log-likelihood ratio test statistic,
% while those corresponding to
$1.94\times10^{-3}$ for the $p$-values $p(\sum X_{mg}^j > x)$ and
$1.87\times10^{-3}$ for the CaMP scores.
%are $1.94\times10^{-3}$ and $1.87\times10^{-3}$ respectively.
The ROC curves estimates suggest that the log-likelihood ratio test
statistic provides best discrimination, though differences are small.

%f3 ###

%that have been proved to control the FDR
%Table~\ref{tab2} answers to our primary questions.
%Which procedures on average performs well under a sampling model that
%reflects the actually observed data?
%Does any of the considered method selects too many false cancer genes?
%And, which methods results too conservative?
%The operating characteristics synthesized in Table~\ref{tab2} answer directly
%to these question. The results can be easily interpreted and
%provides a solid basis for choosing among alternative methods,
%controlling that the average error rate is not undesirably high and
%verifying that possible heuristic selection criteria are appropriate.
%The interpretation of the simulation study is only anchored to the
%modeling assumptions formalized in Section~\ref{sec2}. We emphasizes that the
%proposed validation approach takes substantial advantage from the
%large number of simultaneously considered hypothesis for fitting the
%Bayesian model, and then simulating scenarios by following a
%predictive approach.

%The proposed simulation approach allows us to answer also another
%relevant question: what is the variability of the random proportion of
%false discoveries?
% Figure~\ref{fig2} compares the variability of the considered procedures for
%controlling the false discovery rate.
%Each box plot is representative of the empirical distribution,
%observed by means of simulations, of the proportion of erroneously
%rejected hypothesis.
%The lines in the box plots corresponds to the medians and to the
%$10$-th, $25$-th, $75$-th and $90$-th percentiles of the empirical
%distributions.

%s5 ###
\section{Discussion}\label{sec5}

The investigation of somatic alterations is of primary interest in
cancer research.
Recent sequencing technologies have brought new insight into this
question by revealing a landscape characterized by mutations involving
a large number of driver genes, each altered in a relatively small
fraction of tumors.
When the earlier of these studies began to emerge, the cancer research
community was faced with unexpected heterogeneity and complexity, which
required a significant change of perspective in both basic and clinical
cancer investigations.
The earliest genome-wide investigations [\citet{Sj07}] proposed this
change of landscape based on a~relative small number of samples.
A key element in support of this proposal were estimates of the
statistical significance of the reported list of driver genes [\citet
{Sj07}]. These estimates were challenged: alternative approaches were
proposed which would have led to reporting a drastically reduced number
of candidate drivers at the same significance level.

Subsequent studies have provided strong supporting evidence for this
new landscape, as well as validation for the driver role of many of the
individual genes initially identified in \citet{Sj07}. However, from a
statistical standpoint, it remains very interesting to understand
whether the initial conclusion was statistically sound based on the
evidence available at the time, in part because similar problems will
arise again in other cancer types and in other fields of genomics and
evolutionary biology. Thus, our focus in this article has been the
rigorous evaluation of methods for the identification of driver genes,
with special emphasis on the methods that have been instrumental in the
change of landscape we just described.

In the statistical literature, two common approaches for the evaluation
of methodologies are asymptotic properties and scenario-based
simulation studies. Asymptotic conclusions can be difficult to
extrapolate to small samples. Scenario-based simulations can lack
objectivity and comprehensiveness, and it can be a challenge to gauge
whether the conclusions are applicable or not to a specific context of
interest. To overcome these difficulties, we have proposed and
implemented an alternative concept, which we hope will be very broadly
applicable across statistics, and contribute to a more objective
assessment of alternative methods. The idea is to construct a ``super
partes'' model that is (a) independent of the approaches being compared;
(b) fits the data and available substantive knowledge well; and (c) can
produce artificial data sets accounting for all relevant uncertainties,
including parameter and potentially model uncertainty. This model is
then used to simulate objective data-driven scenarios for method comparison.
In this article our implementation of the super-partes model is based
on Bayesian nonparametrics, an approach that can satisfy all three of
the requirements above.

Strengths of the approach we proposed are the clear interpretation of
both the assumptions captured by the Bayesian model and the average
operating characteristics. The probability model's ability to reproduce
the data structure allows to effectively interpret the
results.
% For example, our model addresses the gene-to-gene dependencies and
%uncertainty in the evaluation of frequentist
%properties of statistical procedures for multiple testing.
The proposed evaluation scheme could be extended further to allow for
more complex assumptions such as
dependency among genes belonging to common functional pathways.

When applied to the controversy surrounding FDR control of early cancer
genome studies, our method shows that the estimates provided in \citet
{Sj07} are quite accurate despite the approximations used.
Also, the \citet{BH95} method is severely conservative. Last, the
Empirical Bayes method and the Storey method based on likelihood ratios
emerge as the preferred choices, though the margin of improvement is
dependent on the control level.
%Our goal, of developing a simple simulation scheme for a better
%interpretation of lists of prioritized genes, is motivated by
%discussions on recent
%sequencing studies. The simplicity of the proposed approach is both
%conceptual and technical: the idea of comparing and assessing methods
%by synthetically reproducing a data set of interest is intuitive and,
%on the other %hand, the choice of a Dirichlet mixture prior allows us
%to rely on a rich literature. Appealing computational and theoretical
%properties of this model made it almost ubiquitous in the
%nonparametric Bayesian literature.
%
%The discussed simulations quantify the conservative behavior of the
%substantial deviations of the Empirical Bayes method from the target
%FDR %when $\alpha$ is set to $20\%$. These and similar considerations
%allows a critical interpretation of lists of putative drivers.

\section*{Acknowledgments}
We thank the Editor and two referees for helpful comments. We thank Ken
Kinzler, Victor Velculescu and Bert Vogelstein for sharing their
invaluable insight on the issues discussed in our analysis.

%suskaldyti doi

% imsref loaded by smiklovaite, 2011-02-18 10:17:03
% imsref loaded by smiklovaite, 2011-02-18 10:28:16
% imsref loaded by smiklovaite, 2011-02-18 10:34:29
% imsref loaded by smiklovaite, 2011-02-18 10:41:37
%

\printaddresses

\end{document}